\documentclass[a4,12pt]{article}
\usepackage{graphicx,psfrag,feynmp,amssymb,cite}

\renewcommand{\thefootnote}{\fnsymbol{footnote}}

\newcommand{\prepr}[1] {\begin{flushright}  {\bf #1} \end{flushright} \vskip 1.cm}
\newcommand{\titul}[1] {\boldmath \begin{center}{\Large {\bf #1 } } \end{center}
\vskip 0.8cm}

\newcommand{\autor}[1] {\begin{center}  {\bf \lineskip .3cm #1  }
                        \end{center} }

\newcommand{\lugar}[1] {\begin{center}  {\normalsize \bf \it #1   } \end{center}}
\newcounter{muni}

\begin{document}
\begin{fmffile}{feynmangraphs}
\hbadness=10000
\pagenumbering{arabic}
\begin{titlepage}

\prepr{hep-ph/0109091\\
\hspace{30mm} KIAS-P01037 \\
\hspace{30mm} KEK-TH-782 \\
\hspace{30mm} September 2001}

\begin{center}
\titul{\bf Direct CP Asymmetry of $B\to X_{d,s}\gamma$ in a model with 
Vector quarks}

\autor{A.G. Akeroyd$^{\mbox{1}}$\footnote{akeroyd@kias.re.kr},
S. Recksiegel$^{\mbox{2}}$\footnote{stefan@post.kek.jp} }
\lugar{ $^{1}$ Korea Institute for Advanced Study,
207-43 Cheongryangri-dong,\\ Dongdaemun-gu,
Seoul 130-012, Korea}

\lugar{ $^{2}$ Theory Group, KEK, Tsukuba, Ibaraki 305-0801, Japan }

\end{center}

\vskip2.0cm

\begin{abstract}
\noindent{We investigate the effect of vector quarks on 
the inclusive decays $B\to X_{d,s}\gamma$. 
We show that the branching ratio of $B\to X_{d}\gamma$ can differ
sizably from the SM and MSSM predictions, being 
enhanced to present experimental observability or suppressed
such that present runs of the B factories would not observe it.
Current measurements of the direct CP asymmetry (${\cal A}_{CP}$)
for $B\to X_{s}\gamma$ are sensitive to the contribution 
from $B\to X_{d}\gamma$. 
For a sufficiently enhanced BR$(B\to X_{d}\gamma$) we show that 
the dominant contribution to the combined asymmetry
may be from $B\to X_{d}\gamma$. Thus any large value for
${\cal A}_{CP}$ should not immediately be attributed to $B\to X_{s}\gamma$,
which stresses the importance of good $K/\pi$ separation.}

\end{abstract}

\end{titlepage}
\thispagestyle{empty}
\newpage

\pagestyle{plain}
\renewcommand{\thefootnote}{\arabic{footnote} }
\setcounter{footnote}{0}

\section{Introduction}

Theoretical studies of rare decays of $b$ quarks have attracted
increasing attention since the start of the physics program at the 
$B$ factories at KEK and SLAC. Both $B$ factories
are running to expectations and in excess of 30 fb$^{-1}$
of data has been accumulated by each experiment. 
The much anticipated measurement of
$\sin2\phi_1$ has established CP violation in the $B$ system 
\cite{Abe:2001xe, Aubert:2001nu}.

Many rare decays will be observed for the first time
over the next few years, and theoretical studies of
such decays in the context of models with new physics will
continue to intensify. In this paper we are concerned 
with the decays $b\to d\gamma$ and $b\to s\gamma$ in a model
with vector (or singlet) quarks. In such a framework the CKM matrix is
necessarily non--unitary, leading to flavour changing neutral currents
at tree--level.

It has been known for a considerable time that 
the inclusive decay $B\to X_s \gamma$ is a sensitive probe
of new physics \cite{Hewett:1994bd}. It has been measured at 
CLEO \cite{Alam:1995aw, Chen:2001fj},
ALEPH \cite{Barate:1998vz} and BELLE \cite{Abe:2001hk}.  
The current branching ratio (BR) is in agreement with the 
Standard Model (SM) expectation \cite{Chetyrkin:1997vx} but  
leaves room for new physics contributions. Preliminary measurements
of the CP asymmetry have been made in the inclusive channel 
$B\to X_s \gamma$ \cite{Coan:2001pu}
and the exclusive channel $B\to K^*\gamma$ \cite{Coan:2000kh},
\cite{Ushiroda:2001sb}. Results with
higher precision are expected to come from the B factories
in the future, especially after a possible luminosity upgrade.

Transitions of the form $b\to d\gamma$ have so far remained 
unobserved and in the SM 
are suppressed relative to $b\to s\gamma$ by a factor
$|V_{td}/V_{ts}|^2\approx 1/20$. Experimental upper limits exist
for the branching ratios (BRs) of the exclusive decay channels, 
$B\to \rho^0\gamma$ and $B\to \rho^+\gamma$. 
CLEO  \cite{Coan:2000kh}
obtains $\le 1.7\times 10^{-5}$ and
$\le 1.3\times 10^{-5}$ respectively, with corresponding measurements
by BELLE \cite{Ushiroda:2001sb} of
$\le 1.06\times 10^{-5}$ and $\le 0.99\times 10^{-5}$. 
 
There is considerable motivation for measuring the 
BR and CP asymmetry (${\cal A}_{CP}$) of the inclusive channel
$B\to X_d\gamma$:

\begin{itemize}

\item[{(i)}] It provides a theoretically clean way of measuring
$V_{td}$, as proposed in \cite{Ali:1992qs, Ali:1998rr}
\item[{(ii)}] ${\cal A}_{CP}$ in the SM is sizeable, and 
much larger than that for $b\to s\gamma$ \cite{Ali:1998rr}.

\item[{(iii)}]$ {\cal A}_{CP}$ is sensitive
to new physics at the weak scale 
\cite{Asatrian:1997kc,Asatrian:1999sg,Asatryan:2000kt,Kiers:2000xy,Akeroyd:2001cy}

\item[{(iv)}] $b\to d\gamma$ transitions sizably affect
the measurements of ${\cal A}_{CP}$ for $b\to s\gamma$ \cite{Coan:2001pu}.
Therefore 
knowledge of ${\cal A}_{CP}$ for $b\to d\gamma$ is essential, in order to
compare experimental data with the theoretical prediction 
in a given model \cite{Akeroyd:2001cy}. 

\item[{(v)}] ${\cal A}_{CP}$ for the 
combined signal of $B\to X_s\gamma$ and 
$B\to X_d\gamma$ is expected to be close to zero in the SM
\cite{Soares:1991te,Kagan:1998bh}, due the
real Wilson coefficients and the unitarity of the CKM matrix. 
Both of these conditions can be relaxed in models beyond the SM.

\end{itemize}

The exclusive decays (e.g. $B\to \rho^0,\omega^0 \gamma$ and $B\to \rho^+\gamma$)
are expected to be observed for the first time
at the B factories. A measurement of the inclusive decay, although
challenging, may be feasible due to the improved $K/\pi$ separation which is
necessary to reduce the large background from $b\to s\gamma$ decays.
The much larger ${\cal A}_{CP}$  of $b\to d\gamma$  with respect
to $b\to s\gamma$ (in the SM) is expected to compensate for 
its smaller branching ratio in terms of experimental observability.

The contribution of vector (i.e.\ $SU(2)$--singlet) quarks 
\cite{delAguila:1986ne,Branco:1993wr} to B physics 
processes has received increasing attention over the last few years
\cite{Gronau:1997rv,Barger:1995dd,Barenboim:2001fd,Handoko:1995xw,
Bhattacharyya:1994vp,Aoki:2000ze,Chang:2000gz,vectorquarkpapers}.
Such particles appear in Grand Unified Theories (GUTs) \cite{Barger:1995dd}
and mix with the ordinary quarks, thus rendering the
CKM matrix non--unitary. We will be working in the context of a 
model with one $U$-- and one $D$--type vector quark. 
The presence of such particles affects the decays
$B\to X_s\gamma$ and $B\to X_d\gamma$. Due to less 
stringent experimental constraints on the VQ parameters, 
the latter decay in particular may have a BR and ${\cal A}_{CP}$ very 
different from that expected in the SM and the Minimal Supersymmetric
Standard Model (MSSM).

Our work is organised as follows: In section 2 we introduce
our formalism and the vector quark model (VQ--model). In section 3 we outline 
our approach to calculate the VQ--contributions to
the BRs and ${\cal A}_{CP}$ of $b\to d\gamma$ and $b\to s\gamma$.
Section 4 presents the numerical results 
and section 5 contains our conclusions.

\boldmath
\section{Vector quarks and the decays $b\to {s,\!d}\,\gamma$}
\unboldmath

There is much theoretical and experimental motivation to
study the ratio
\begin{equation}
R={BR(B\to X_d\gamma)\over {BR(B\to X_s\gamma)}}
\end{equation}
because it provides a clean handle on the ratio
$|V_{td}/V_{ts}|^2$ \cite{Ali:1998rr}.
In the context of the SM, $R$ is expected to be in the 
range $0.017 < R < 0.074$, corresponding
to BR$(B\to X_d\gamma)$ of order $10^{-5}$. 
$R$ stays confined to this range 
in many popular models beyond the SM. This is because new 
particles such as charginos and charged Higgs bosons in the MSSM
contribute to $b\to s(d)\gamma$ 
with the same CKM factors. Therefore $C_7$ is universal to both
decays and cancels out in the ratio $R$. 
In a model with vector quarks this is not the case, and we 
shall see that $R$ can be suppressed or enhanced 
with respect to the SM\footnote{SUSY models with 
non--flavour diagonal SUSY--breaking terms can also
suppress or enhance $R$ \cite{Gabbiani:1996hi,Arhrib:2001gr}}.

The dominant source of CP--Asymmetry in the decay modes
$B\to X_{s/d}\gamma$ is direct CP--violation. Mixing induced
CP--violation is strongly suppressed by a factor of $m_{s/d}/m_b$.
The interference between mixing and decay necessary for this
kind of CP--violation can only occur between identical final
states. The photons from $b\to s/d\,\gamma$ are predominantly
left--handed, while those from $\bar b\to \bar s/\bar d\,\gamma$ 
are predominantly right--handed, so interference is possible only
between strongly suppressed contributions. (Since the weak 
interaction is left--handed, the spin flip in the quark line 
in the penguin loop must occur on an external leg, and this is 
proportional to the mass of the external quark involved.) 

Mixing induced CP--violation becomes important for 
$B\to X_{s/d}\gamma$ only if new, right--handed weak interactions
are possible, because then the spin flip can occur inside the loop
\cite{Atwood:1997zr}. This is not relevant for the model we consider
because our Vector Quarks are $SU(2)$--singlets and do not introduce
right--handed couplings to the weak bosons.

The direct CP--Asymmetry is given by
\begin{equation}
{\cal A}^{d\gamma(s\gamma)}_{CP}={{\Gamma(\overline B\to X_{d(s)}\gamma)-
\Gamma(B\to X_{\overline d(\overline s)}\gamma)}
\over {\Gamma(\overline B\to X_{d(s)}\gamma)+\Gamma(B\to X_{\overline
d(\overline s)}\gamma)}}={\Delta\Gamma_{d(s)}\over \Gamma^{tot}_{d(s)}}
\label{ACPdef} \end{equation}
In the SM ${\cal A}_{CP}^{d\gamma}$ is expected to lie in the 
range $-7\%\le {\cal A}_{CP}^{d\gamma}\le -35\%$ 
\cite{Ali:1998rr}, where the uncertainty arises from 
varying the Wolfenstein parameters 
$\rho$ and $\eta$ in their allowed ranges. 
Also included is the dependence of ${\cal A}_{CP}^{d\gamma}$
on the scale $\mu_b$ which arises from varying
$m_b/2\le \mu_b \le 2m_b$. For definiteness we fix
$\mu_b=4.8$ GeV, and find $-5\%\le A_{CP}^{d\gamma}\le -28\%$.
Therefore  ${\cal A}_{CP}^{d\gamma}$ is much larger than
${\cal A}_{CP}^{s\gamma}$ ($\lesssim 0.6\%$). 

If $b\to d\gamma$ and $b\to s\gamma$ 
cannot be properly separated, then only $A_{CP}$ of
a combined sample can be 
measured.\footnote{This is also true (though much less
 important numerically) for the total rate, where the
 $b\to d\gamma$ contribution should be subtracted from the
 ``$b\to s\gamma$'' sample as done e.g. in \cite{Chen:2001fj}}
Neglecting the masses
$m_s$ and $m_d$ and subleading CKM--factors
$V_{us(d)}^*V_{ub}$ against $V_{ts(d)}^*V_{tb}$
(i.e.\ considering only the leading terms),
it has been shown \cite{Soares:1991te,Kagan:1998bh}
that ${\cal A}_{CP}^{s\gamma}$ and
$A_{CP}^{d\gamma}$ cancel each other\footnote{An improved
analysis with non-vanishing quark masses showed that this 
cancellation still holds to a very high degree for both the 
inclusive and exclusive decays\cite{Hurth:2001yb}}.

In the presence of new physics such a cancellation does not
occur, as was shown in \cite{Akeroyd:2001cy} in the context of
the effective SUSY model.
As stressed in \cite{Akeroyd:2001cy}, a 
reliable prediction of 
${\cal A}^{d\gamma}_{CP}$ in a given model is necessary 
since it contributes to the measurement of ${\cal A}^{s\gamma}_{CP}$.
The CLEO result \cite{Coan:2001pu} is sensitive to a weighted sum of CP asymmetries, 
given by:
\begin{equation}
{\cal A}^{exp}_{CP}=0.965{\cal A}^{s\gamma}_{CP}+0.02{\cal A}^{d\gamma}_{CP}
\label{CLEOeq} \end{equation}  
The latest measurement stands at $-27\% < {\cal A}^{exp}_{CP} < 10\%$ 
(90\% C.L.) \cite{Coan:2001pu}. 
The small coefficient of ${\cal A}^{d\gamma}_{CP}$ is caused by
the smaller BR$(B\to X_d\gamma)$ (assumed to be $1/20$ that
of BR$(B\to X_s\gamma)$) and inferior detection efficiencies.

If the detection efficiencies for both decays were identical,
this measured quantity would coincide with the weighted sum of the asymmetries
\begin{equation}
{\cal A}_{CP}^{s\gamma+d\gamma} = { {\rm BR}^{s\gamma} {\cal A}_{CP}^{s\gamma}
 +{\rm BR}^{d\gamma} {\cal A}_{CP}^{d\gamma} \over {\rm BR}^{s\gamma} +{\rm BR}^{d\gamma}}\,.
\label{ACPcombined} \end{equation}  

The two terms in eqs.~(\ref{CLEOeq},\ref{ACPcombined}) can
be of equal or of opposite sign, i.e.\ they can contribute 
constructively or destructively to the combined asymmetry.
Since BR$(B\to X_d\gamma)$ may be enhanced in the
VQ--model, the relative strength of its contribution to 
${\cal A}_{CP}^{s\gamma+d\gamma}$
may be increased. The non--negligible contribution of
$b\to d\gamma$ to this combined asymmetry 
should be verifiable at proposed future high luminosity  
runs of $B$ factories.
For integrated luminosities of 200 fb$^{-1}$ (2500 fb$^{-1}$), \cite{alex}
anticipates a precision of $3\%\,(1\%)$ in the measurement of
${\cal A}^{exp}_{CP}$. 

We shall be working in a model with an extra generation of vector
quarks, $U$ and $D$. Both $U$ and $D$ are singlets under $SU(2)$ and
in the interaction basis (denoted by $U'$ and $D'$) do not couple to
the $W$. Their mass is generated by terms of the form:
\begin{equation}
-f^{i4}_d\overline \psi_L^iD'_R\phi-f^{i4}_u\overline \psi_L^iU'_R\overline 
\phi+{\rm h.c.}\,,
\end{equation}
where $f^{i4}$ ($i=1\!\to\! 3$) are Yukawa couplings and $\overline
\psi_L^i=(\bar u_i,\bar d_i)_L$.
Thus $U'$ ($D'$) will mix with the up (down) type quarks, resulting
in the (undiscovered) mass eigenstates $U$ and $D$. The known 
quarks hence contain small amounts of the $U'$ and $D'$ weak 
eigenstates. A feature of the vector quark model is the extended
$4\times 4$ CKM matrix which is now not unitary. This can be
clearly seen from the charged current interaction in the
interaction basis:
\begin{equation}
{\cal L}={g\over \sqrt 2}(W^-_\mu J^{\mu+}+W^+_\mu J^{\mu-})
\end{equation}
where\footnote{It is obvious where $u_L$ denotes the 4--vector of up--type quarks and
where it denotes the up--quark only}
\begin{equation}
J^{\mu-}=\overline u_L'a_W \gamma^\mu d_L'=\overline u_L V_{\rm CKM} \gamma^\mu d_L
\qquad {\rm and} \qquad
u_L=\left(\begin{array}{c}u_L\\ c_L \\ t_L\\ U_L \end{array}\right) \!,\,\,
d_L=\left(\begin{array}{c}d_L\\ s_L \\ b_L\\ D_L \end{array}\right) 
\end{equation}

Here the matrix $a_W$ is Diag(1,1,1,0), reflecting the fact that
the charged current couplings to the known quarks are flavour diagonal, while
the vertex $W$-$U'$-$D'$ is absent. In the mass basis
the vertices $W$-$U$-$d$ ($d$ denoting any down--type quark) are generated 
and thus an extra diagram with an internal $U$ quark contributes to the 
decay $b\to s\gamma,d\gamma$. The $4\times 4$ CKM matrix is then simply given by 
\begin{equation}
V_{\rm CKM}=U^{u\dagger}_L a_W U^{d}_L
\end{equation}
where $U^{u\dagger}_L$ and $U^{d}_L$ rotate the mass eigenstates
to the interaction eigenstates. Clearly $V_{\rm CKM}$ is not unitary due 
to the presence of $a_W$. A consequence of this is the generation
of FCNC vertices, where the effective coupling for the vertex
$Z$-$d_i$-$d_j$ is $(V^\dagger V)_{ij}$. This gives rise to a
further class of diagrams that contributes to $b\to s\gamma,d\gamma$
(c.f.\ section 3).

These FCNC vertices are strongly constrained by current experiments.
The vertices $Z$-$b$-$s$ and $Z$-$b$-$d$ are e.g.\ constrained by
the non--observance of $B\to X_sl^+l^-$ and $B\to X_dl^+l^-$,
respectively \cite{Barenboim:2001fd}.
(Note that the first evidence for $B\to K\mu^+\mu^-$ has recently 
been reported by the BELLE collaboration \cite{Abe:2001qh}.)

We now briefly summarise the previous works which considered  
the effect of vector quarks on $B \to X_{d(s)} \gamma$.
(Work on these decays in other models has already been summarised 
in \cite{Akeroyd:2001cy}.)

\cite{Handoko:1995xw,Bhattacharyya:1994vp}
studied a model with a $D$ type vector quark which induces $Z$ and $h$
FCNC contributions to $B \to X_{d(s)} \gamma$. It was shown that
these contributions have negligible impact on $B \to X_{s} \gamma$,
but can enhance or suppress $R$. 
Implications for ${\cal A}^{s\gamma}_{CP}$ were found to be small.

The effect of the $U$-$W$ contribution on ${\cal A}^{s\gamma}_{CP}$
was considered in \cite{Kagan:1998bh}. 
In a model with both $U$ and $D$ vector quarks (where the 
$D$--contribution could actually be neglected) \cite{Aoki:2000ze},
the $U$-$W$ mediated contribution to BR$(B \to X_{s} \gamma$) was shown
to permit BRs anywhere within the experimental limits. 
More accurate measurements of BR($B \to X_{s} \gamma$) 
at the B factories will further restrict the available
parameter space for VQ--models.

To our knowledge the combined effect of $U$ and $D$ vector quarks
on BR$(B \to X_{d} \gamma$) has not yet been considered. 
In addition, no analysis of ${\cal A}^{d\gamma}_{CP}$ 
in any model with vector quarks has been carried out. 
Given the importance of the measurement of the
combined asymmetry ${\cal A}^{s\gamma+d\gamma}_{CP}$, 
we will also study the correlation of the individual asymmetries
${\cal A}^{s\gamma}_{CP}$ and ${\cal A}^{d\gamma}_{CP}$ 
as a function of the VQ parameters. 

\boldmath
\section{Direct CP Asymmetry in $B \to X_{d,s} \gamma$}
\unboldmath

\begin{figure}
\begin{center}
\begin{fmfgraph*}(220,100)
\fmfleft{i1,i2,i3} \fmfright{o1,o2,o3}
\fmf{phantom}{i1,v2}
\fmf{phantom,tension=0.4}{v2,v3}
\fmf{phantom}{v3,o1}
\fmffreeze
\fmf{fermion,label=$b$}{i1,v2} 
\fmf{fermion,label.side=right,label=$s,,d$}{v3,o1} 
\fmf{fermion,label=$u,, c,, t,, U$}{v2,v3}
\fmf{photon,label.side=right,label=$\gamma$,tension=1.5}{v1,o3}
\fmf{photon,label=$W$,left=.7,label.dist=8}{v2,v1} 
\fmf{photon,left=.2}{v1,v3} 
\fmfdot{v1,v2,v3}
\end{fmfgraph*}
\hspace{1cm}
\begin{fmfgraph*}(220,100)
\fmfleft{i1,i2,i3} \fmfright{o1,o2,o3}
\fmf{phantom}{i1,v2}
\fmf{phantom,tension=0.4}{v2,v3}
\fmf{phantom}{v3,o1}
\fmffreeze
\fmf{fermion,label=$b$}{i1,v2} 
\fmf{fermion,label.side=right,label=$s,,d$}{v3,o1} 
\fmf{fermion}{v2,v0,v3}
\fmffreeze
\fmf{phantom,label=$d,, s,, b,, D$}{v2,v3}
\fmf{photon,label.side=right,label=$\gamma$,tension=1.5}{v0,o3}
\fmf{photon,label=$Z,, H$,left=1.15,label.dist=8}{v2,v3} 
\fmfdot{v2,v3,v0}
\end{fmfgraph*}
\end{center}
\caption{Diagrams contributing to $b\to s,\!d \,\gamma$ in the VQ--model
\label{diagrams}}
\end{figure}
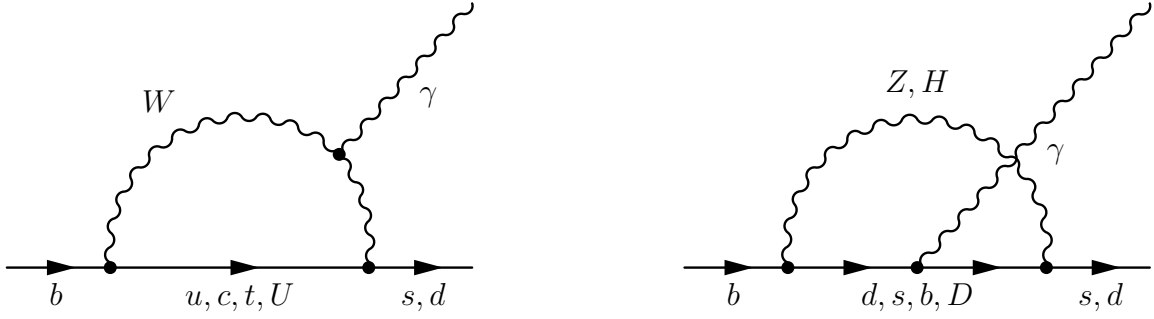

In this section we explore the effect of $U$ and $D$--type 
vector quarks on both BR$(B \to X_{d,s} \gamma$)
and the direct CP asymmetries, ${\cal A}^{d,s\,\gamma}_{CP}$.
Diagrams contributing to $b\to s,\!d \,\gamma$ in the VQ--model are
shown in fig.~\ref{diagrams}.

The effective  Hamiltonian for $b \to d \gamma$ is given by
\begin{equation}
{\cal H}_{eff} = -{4 G_F \over \sqrt{2}} V_{td}^{*} V_{tb} \sum_{i=1}^{8}
C_{i}(\mu_b) Q_{i}(\mu_b)
\label{Heff}
\end{equation}
where $Q_{i}(\mu_b)$ are the current density operators for the $\Delta B =1$
transitions and $C_{i}(\mu_b)$ are their Wilson coefficients.
The relevant operators for $b \to d \gamma$ decay are given by
\begin{eqnarray}
Q_{2} &=& \bar{d}_{L} \gamma^{\mu}c_{L} \bar{c}_{L} \gamma^{\mu} b_{L}, 
\nonumber \\
Q_{7} &=& {e \over 16 \pi^2} m_b \bar{d}_{L} \sigma^{\mu\nu}b_{R}F_{\mu\nu},
\label{op} \\
Q_{8} &=& {g_s \over 16 \pi^2} m_b \bar{d}_{L} \sigma^{\mu\nu} T^{a}
b_{R}G^{a}_{\mu\nu}. \nonumber 
\end{eqnarray}
The analogous formulae for the $b \to s\gamma$ decay can obtained 
from Eqs.(\ref{Heff}) and (\ref{op}) by making the
replacement $d\to s$.

We use formula (27) from \cite{Kagan:1998bh}:
\begin{eqnarray}
{\cal A}_{CP}^{d(s)\gamma}&=&{\alpha_s(m_b)\over |C_7|^2}
\Big\{{40\over 81} {\rm Im}[(1+\Delta_{d(s)}) C_2 C_7^*]-{4\over 9}
{\rm Im}[C_8 C_7^*]+{8z\over 9} \left(v(z)+b(z,\delta)\right)
\\
&& \hspace{10mm}  \times {\rm Im}[(1+\epsilon_{d(s)}+\Delta_{d(s)})C_2 C_7^*]
+{8z\over 27}\,b(z,\delta) \,{\rm Im}
[(1+\epsilon_{d(s)}+\Delta_{d(s)})C_2 C_8^*]\Big\} \,,\nonumber 
\label{Acp}
\end{eqnarray}
where in our notation
$\Delta_x=(z_{xb}-V_{Ux}^* V_{Ub})/(V_{tx}^* V_{tb})$ for ($x=d,s$).

The expressions for the
Wilson coefficients at the $M_W$ scale may be divided into the 
charged current and neutral current mediated contributions, where
only the former exist in the SM. Contrary to the SM case,
the charged current mediated contribution contains a term stemming
from a loop with the $U$ vector quark. 

Thus at the $M_W$ scale we have:
\begin{equation}
C_7=C_7^{W}+C_7^{ZH}
\end{equation}
We will use the leading order expressions for $C_7$ (with the
constant terms included in the Inami--Lim functions) which
can be found in \cite{Chang:2000gz}.

In the SM the CKM--factors  are customarily included in the 
pre-factor of the effective Hamiltonian. This is possible
because the top--quark loop is the dominant contribution to
$C_7$ and the CKM--factor $V_{cx}^* V_{cb}$ in $C_2$ 
can be reexpressed in terms of $V_{tx}^* V_{tb}$ utilizing 
CKM unitarity. 

In the vector quark model this is not possible anymore.
For $C_7$, the constant terms in the Inami--Lim functions that 
are cancelled in the SM due to CKM unitarity, have to be taken 
into account. Also in replacing $V_{cx}^* V_{cb}$ in $C_2$,
additional terms arising from CKM non--unitarity are generated.
This results in different Wilson coefficients for $b\to s\gamma$ 
and $b\to d\gamma$.

For comparison with standard calculations,
we keep the term $V_{td(s)}^{*} V_{tb}$ in the pre-factor of 
${\cal H}_{eff}$ and therefore have to divide all but the
top--quark contributions to the Wilson coefficients by this factor.

We define $C^{s\gamma}_7$ and $C^{d\gamma}_7$
as the Wilson coefficient for  $b\to s\gamma$ and $b\to d\gamma$
respectively.
The magnitude of $|C^{s\gamma}_7|$ is constrained by measurements 
of BR($B\to X_s\gamma$) \cite{Aoki:2000ze}. 

In contrast, $C^{d\gamma}_7$ is only very weakly constrained
due to the non--observation of $b\to d\gamma$.
Recent direct searches for the exclusive decay
$B\to \rho \gamma$ give an improved upper bound on $R_{\rm excl}$ 
for exclusive decays of $R_{\rm excl}\lesssim 0.19$ (90\% C.L.)
\cite{Ushiroda:2001sb}.
One expects a weaker bound on the inclusive $R$, since
estimates of $R_{\rm excl}/R$ are smaller than 1
\cite{Bhattacharyya:1994vp}.

On the other hand the presence of $b\to d\gamma$ events
in the samples of the inclusive measurements of 
BR($B\to X_s\gamma$) \cite{Chen:2001fj}
also constrains $R$. This bound, however, is not relevant for
our analysis, since in the vector quark model the prediction
for $b\to s\gamma$ can easily be lowered to account for a
higher admixture of $b\to d\gamma$ in the experimental sample.

\section{Numerical Results}

We vary the vector quark parameters in the ranges given in
table \ref{parameters}. The Wolfenstein parameters $\rho$
and $\eta$ quantify the current uncertainty in the CKM
parameters related to the mixing of the first and the third
generation of quarks. Our results are virtually independent
of our choice of $m_{U/D}$ and $M_H$. 

In the $s$--sector we use $|V_{Us}^*V_{Ub}|<0.004$ which is
1/3 smaller than the bound derived from BR$^{s\gamma}$
in \cite{Aoki:2000ze}, we take this smaller value in the light of
recently improved measurements of $b\to s\gamma$.
The non--unitarity parameter $z_{sb}$ ($z_{db}$) is constrained 
from non--observation of $B\to X_s \ell^+\ell^-$
($B\to X_d \ell^+\ell^-$) \cite{Barenboim:2001fd}.

Our results depend crucially on the magnitude of
$|V_{Ud}^*V_{Ub}|$. Bounds from CKM unitarity on this
quantity are rather weak (of the order of the
uncertainty in the CKM parameters in the $d$--sector)
and $B\to X_d\gamma$ has not
yet been observed, so the only restrictions on this
parameter come from $B^0 \bar B^0$--mixing 
\cite{Barger:1995dd,Branco:1993wr,Gronau:1997rv}.
The exact bounds on $|V_{Ud}^*V_{Ub}|$ from 
$B^0 \bar B^0$--mixing are not clear and depend on 
the tree level $Z$--mediated contributions
as well as the rather large uncertainties in the SM
prediction of $B^0 \bar B^0$--mixing (bag parameter, etc.).
We give results for different choices of $|V_{Ud}^*V_{Ub}|$
up to 0.01 ($\approx |V_{td}^*V_{tb}|$).
A more detailed study of $B^0 \bar B^0$--mixing
in the VQ model will be presented elsewhere \cite{newakeroydreck}.

\begin{figure}
\begin{center}
\psfrag{XXX}{${\cal A}^{\rm s\gamma}_{CP}$}  \psfrag{YYY}
 {${\cal A}^{\rm d\gamma}_{CP}$}
\includegraphics[width=10cm]{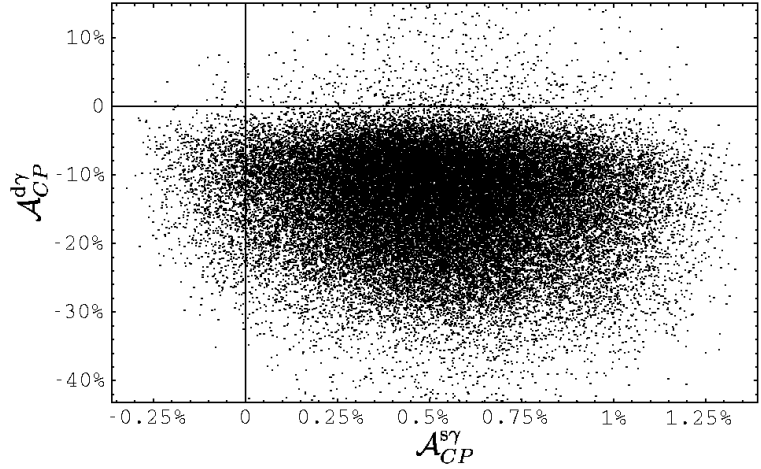}
\end{center}
\vspace*{-.5cm}
\caption{CP asymmetry of $b\to d\gamma$ against
CP asymmetry of $b\to s\gamma$. The VQ--parameters
are varied in the ranges given in table \ref{parameters};
each point in the diagram corresponds to one random point in
parameter space.}
\label{scatter}
\end{figure}

In figure \ref{scatter} we plot ${\cal A}^{\rm d\gamma}_{CP}$
against ${\cal A}^{\rm s\gamma}_{CP}$.
\begin{table}
\begin{tabular}{|c||c|c|c|c|c|c|c|c|c|c|}
\hline
figure\#&$\rho$&$\eta$&$m_{U/D}$&$M_H$&$|V_{Us}^*V_{Ub}|$&$|V_{Ud}^*V_{Ub}|$&$|z_{sb}|$&
 Arg $z_{sb}\!$&$|z_{db}|$&Arg $z_{db}\!$\\
\hline\hline
2\,(\,min)&-0.1&0.2&250&100&0&0&0&0&0&0\\
2\,(max)&0.4&0.5&1000&200&0.004&0.008&$8.1\!\cdot\! 10^{-4}$&$2\pi$&0.001&$2\pi$ \\
\hline
3--6&0.4&0.5&500&100&0.004&variable&$8.1\!\cdot\! 10^{-4}$&0&0.001&$\pi$\\
\hline
\end{tabular}\\
\caption{Model parameter ranges}
\label{parameters}
\end{table}
It can be seen that while ${\cal A}^{\rm s\gamma}_{CP}$
does not substantially differ from its SM value, 
${\cal A}^{\rm d\gamma}_{CP}$ can vary over a much larger
range.  We restrict ourselves to points in parameter space 
where $|{\cal A}^{\rm d\gamma}_{CP}|<45 \%$.
Asymmetries greater than $50\%$ are attainable, but these
correspond to cases with a virtually
complete cancellation of the SM contribution by the
new VQ contribution ($|C_7^{d\gamma}|\ll 0.3$). Such large
asymmetries are untrustworthy, since our formula for the CP 
asymmetry (\ref{ACPdef}) breaks down in these cases.

\begin{figure}
\begin{center}
\psfrag{XXX}{BR$(b\to d\gamma)\,[10^{-5}]$}  \psfrag{YYY}
 {${\cal A}^{b\to d\gamma}_{CP}$} \psfrag{ZZZ} {$|V_{Ud}^*V_{Ub}|$} 
\includegraphics[width=10cm]{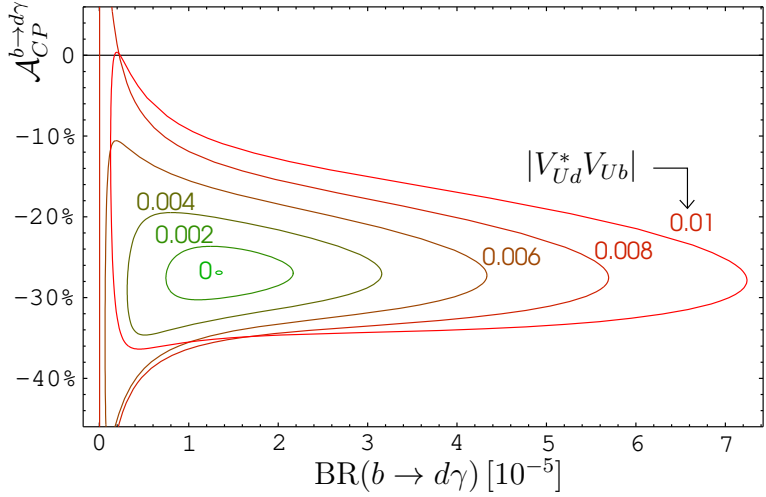}
\end{center}
\vspace*{-.5cm}
\caption{Contour plot of CP asymmetry against branching rate for
$b\to d\gamma$. Contours for different values of $|V_{Ud}^*V_{Ub}|$
are given, and Arg $V_{Ud}^*V_{Ub}$ is varied along the contours}
\label{contours}
\end{figure}

The correlation between ${\cal A}^{\rm d\gamma}_{CP}$ and
BR$^{d\gamma}$ is studied in detail in figure \ref{contours},
where it can be seen that $|{\cal A}^{\rm d\gamma}_{CP}| >45 \%$
occurs only for BR$^{d\gamma} < 10^{-6}$. Branching ratios
of this magnitude would require $\gg 10^8$ $b\bar b$ pairs
to be detected which is beyond the discovery potential of
current $B$ factories. Note that figure \ref{contours}
was obtained with fixed values for the CKM parameters
$\rho$ and $\eta$ as given in table \ref{parameters}. Varying
$\rho$ and $\eta$ shifts the contours in figure \ref{contours}
vertically by the amount of uncertainty in the SM prediction
of ${\cal A}^{\rm d\gamma}_{CP}$.

\begin{figure}
\begin{center}
\psfrag{XXX}{Arg $V_{Ud}^*V_{Ub}\,\,[\pi]$}  \psfrag{YYY}
 {$R={\rm BR}^{d\gamma}/{\rm BR}^{s\gamma}$} \psfrag{ZZZ} {$|V_{Ud}^*V_{Ub}|$}
\includegraphics[width=10cm]{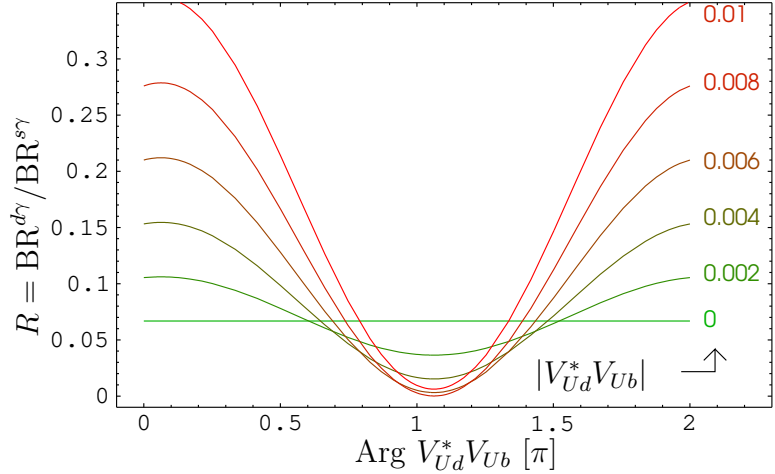}
\end{center}
\vspace*{-.5cm}
\caption{$R$ against Arg $V_{Ud}^*V_{Ub}$
for different values of $|V_{Ud}^*V_{Ub}|$}
\label{valley}
\end{figure}

In figure \ref{valley} we plot the ratio of the branching
ratios for $b\to d\gamma$ and $b\to s\gamma$ against the 
argument of $V_{Ud}^*V_{Ub}$, drawing curves for different 
values of $|V_{Ud}^*V_{Ub}|$.
For real, positive $V_{Ud}^*V_{Ub}$, there is constructive
interference with the SM contribution proportional to
$V_{td}^*V_{tb}=A\lambda^3(1-\rho-i\eta)+{\cal O}(\lambda^4)$.
For real, negative $V_{Ud}^*V_{Ub}$ (corresponding to points
around Arg $(V_{Ud}^*V_{Ub})=\pi$) the interference is
destructive. Note that the smallest values for $R$ are not
obtained for the largest $|V_{Ud}^*V_{Ub}|$, because if
$|V_{Ud}^*V_{Ub}|>|V_{td}^*V_{tb}|$, exact cancellation
cannot occur.

\begin{figure}
\begin{center}
\psfrag{XXX}{Arg $V_{Ud}^*V_{Ub}\,\,[\pi]$}  \psfrag{YYY}
 {${\cal A}^{s\gamma+d\gamma}_{CP}$} \psfrag{ZZZ} {$|V_{Ud}^*V_{Ub}|$}
\includegraphics[width=10cm]{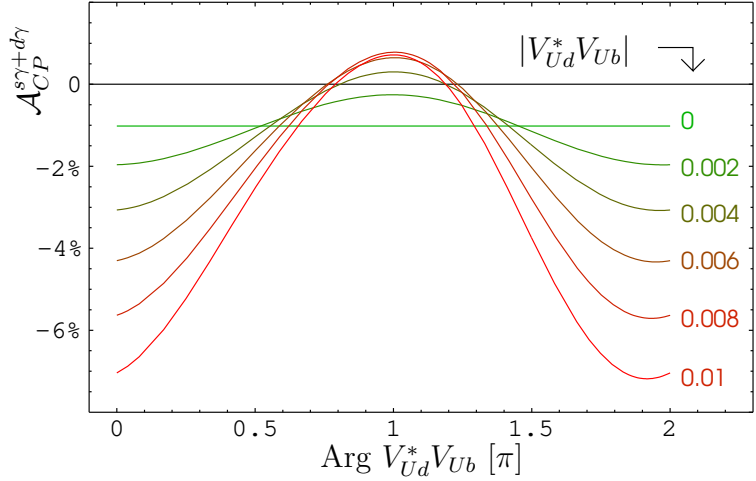}
\end{center}
\vspace*{-.5cm}
\caption{Combined Asymmetry against Arg $V_{Ud}^*V_{Ub}$
for different values of $|V_{Ud}^*V_{Ub}|$}
\label{hill}
\end{figure}

In figure \ref{hill} we plot the combined CP asymmetry as defined
in equation (\ref{ACPcombined}) against the argument of $V_{Ud}^*V_{Ub}$.
We plot several curves for different values of $|V_{Ud}^*V_{Ub}|$ just
like in figures \ref{contours} and \ref{valley}. The extreme values for 
the combined asymmetry are obtained where $R$ is maximal and not --- as 
could be na\"{\i}vely expected --- where the CP asymmetry of $b\to d\gamma$ becomes
much larger than its SM value. This is because large CP asymmetries
in $b\to d\gamma$ as seen in figure \ref{contours} always imply
small values for $R$ which makes these points in parameter space
unimportant for the combined asymmetry.

Note that in our analysis BR$^{s\gamma}$ and 
${\cal A}^{b\to s\gamma}_{CP}$ are close to their SM values.
The huge variations in ${\cal A}^{s\gamma+d\gamma}_{CP}$
stem from the variation in BR$^{d\gamma}$. In wide ranges
of our parameter space, $b\to d\gamma$ actually dominates
the combined asymmetry ! Any large signal observed in
${\cal A}^{\rm s\gamma+d\gamma}_{CP}$ (which is the experimentally
relevant quantity) should not necessarily be attributed
to $b\to s\gamma$ only. This motivates excellent $K$--$\pi$ separation
which is crucial in distinguishing the two decay modes.

\begin{figure}
\begin{center}
\psfrag{XXX}{Arg $V_{Ud}^*V_{Ub} \,\,[\pi]$}  \psfrag{YYY}
 {$\Omega_{d(s)\gamma}\,\,  [10^{-6}]$} \psfrag{ZZZ} {${\cal A}^{b\to d\gamma}_{CP}$} 
\psfrag{AAA}{$\Omega_{d\gamma}^{SM}$}  
\psfrag{BBB}{$\Omega_{s\gamma}^{SM}$}
\psfrag{CCC}{$\Omega_{d\gamma}$} 
\includegraphics[width=10cm]{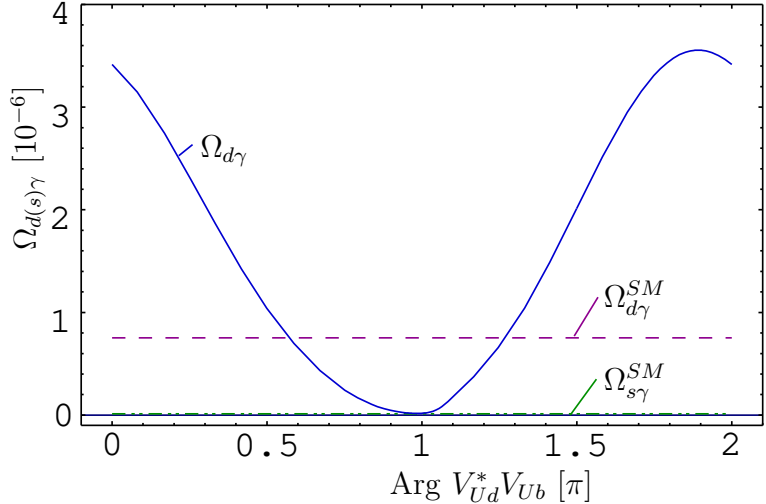}
\end{center}
\vspace*{-.5cm}
\caption{CP asymmetry and observability $\Omega$ against Arg $V_{Ud}^*V_{Ub}$}
\label{observab}
\end{figure}

To obtain a rough estimate on the number of $b\bar b$ pairs $N_{b\bar b}$
required to establish CP violation to a given significance,
following \cite{Soares:1991te} we define the {\it observability}
$\Omega_{d(s)\gamma}=({\cal A}^{b\to d(s)\gamma}_{CP})^2 \cdot {\rm
BR}^{d(s)\gamma}$, where $\Omega_{d(s)\gamma} \propto 1/N_{b\bar b}$.
This observability is plotted in figure \ref{observab} against
Arg $V_{Ud}^*V_{Ub}$, where the horizontal lines indicate the
SM observabilities. (We do not take into account different detection
efficiencies for the two channels.) Despite the suppressed BR$^{d\gamma}$
in the SM, ${\cal A}^{b\to d\gamma}_{CP}$ is much more observable
than ${\cal A}^{b\to s\gamma}_{CP}$.
In most of the parameter space, ${\cal A}^{b\to d\gamma}_{CP}$ has a
higher observability than in the SM. The optimal observability
occurs for the largest values of $R$ (compare fig.~\ref{valley}).

\newpage

\section{Conclusions}
We have studied the effect of a single generation of vector quarks (VQ) 
on the short--distance component of the inclusive decays 
$B\to X_{d,s}\gamma$. We found that ${\cal A}^{b\to s\gamma}_{CP}$ is
relatively insensitive to VQ interactions if the experimental
bounds on BR$^{s\gamma}$ are respected. For the decay mode $b\to d\gamma$
both the decay rate and the CP asymmetry can be significantly different
from their SM values. Phenomenologically most salient is the fact
that the decay rate can be increased up to current experimental observability
or decreased beyond the sensitivity of the current $B$ factories. This
variation of BR$^{d\gamma}$ has a profound impact on the experimentally
relevant CP asymmetry of a combined sample of $b\to s,\!d\,\gamma$ which 
can actually be dominated by ${\cal A}^{b\to d\gamma}_{CP}$. Observation 
of any ${\cal A}^{s\gamma+d\gamma}_{CP}$ significantly different from zero
is a clear sign of new physics, but the theoretical interpretation
should await experimental separation of the individual channels.

\vspace{20mm}
\begin{center}
{\large\bf  Acknowledgements} 
\end{center}

The authors wish to thank A.~Arhrib and Y.~Okada for
useful discussions and comments.
S.R was supported by the Japan Society for the Promotion
of Science (JSPS).

\renewcommand{\theequation}{B.\arabic{equation}}
\setcounter{equation}{0}

\end{fmffile}

\newpage

\begin{thebibliography}{99}

\bibitem{Abe:2001xe}
K.~Abe {\it et al.}  [Belle Collaboration],
Phys.\ Rev.\ Lett.\  {\bf 87}, 091802 (2001).

\bibitem{Aubert:2001nu}
B.~Aubert {\it et al.}  [BaBar Collaboration],
Phys.\ Rev.\ Lett.\  {\bf 87}, 091801 (2001).

\bibitem{Hewett:1994bd}
J.~L.~Hewett,
hep-ph/9406302.

\bibitem{Alam:1995aw}
M.~S.~Alam {\it et al.}  [CLEO Collaboration],
Phys.\ Rev.\ Lett.\  {\bf 74}, 2885 (1995);
S.~Ahmed {\it et al.}  [CLEO Collaboration],
hep-ex/9908022.

\bibitem{Chen:2001fj}
S.~Chen {\it et al.}  [CLEO Collaboration],
hep-ex/0108032.

\bibitem{Barate:1998vz}
R.~Barate {\it et al.}  [ALEPH Collaboration],
Phys.\ Lett.\ B {\bf 429}, 169 (1998).

\bibitem{Abe:2001hk}
K.~Abe {\it et al.}  [Belle Collaboration],
Phys.\ Lett.\ B {\bf 511}, 151 (2001).

\bibitem{Chetyrkin:1997vx}
K.~Chetyrkin, M.~Misiak and M.~Munz,
Phys.\ Lett.\ B {\bf 400}, 206 (1997)
[Erratum-ibid.\ B {\bf 425}, 414 (1997)];
P.~Gambino and M.~Misiak,
Nucl.\ Phys.\ B {\bf 611}, 338 (2001).


\bibitem{Coan:2001pu}
T.~E.~Coan {\it et al.}  [CLEO Collaboration],
Phys.\ Rev.\ Lett.\  {\bf 86}, 5661 (2001).

\bibitem{Coan:2000kh}
T.~E.~Coan {\it et al.}  [CLEO Collaboration],
Phys.\ Rev.\ Lett.\  {\bf 84}, 5283 (2000).

\bibitem{Ushiroda:2001sb}
Y.~Ushiroda  [Belle collaboration],
hep-ex/0104045.

\bibitem{Ali:1992qs}
A.~Ali and C.~Greub,
Phys.\ Lett.\ B {\bf 287}, 191 (1992).

\bibitem{Ali:1998rr}
A.~Ali, H.~Asatrian and C.~Greub,
Phys.\ Lett.\ B {\bf 429}, 87 (1998).


\bibitem{Asatrian:1997kc}
H.~M.~Asatrian, G.~K.~Egiian and A.~N.~Ioannisian,
Phys.\ Lett.\ B {\bf 399}, 303 (1997).


\bibitem{Asatrian:1999sg}
H.~H.~Asatrian and H.~M.~Asatrian,
Phys.\ Lett.\ B {\bf 460}, 148 (1999).

\bibitem{Asatryan:2000kt}
H.~H.~Asatryan, H.~M.~Asatrian, G.~K.~Yeghiyan and G.~K.~Savvidy,
Int.\ J.\ Mod.\ Phys.\ A {\bf 16}, 3805 (2001).

\bibitem{Kiers:2000xy}
K.~Kiers, A.~Soni and G.~Wu,
Phys.\ Rev.\ D {\bf 62}, 116004 (2000).

\bibitem{Akeroyd:2001cy}
A.~G.~Akeroyd, Y.~Y.~Keum and S.~Recksiegel,
Phys.\ Lett.\ B {\bf 507}, 252 (2001).

\bibitem{Soares:1991te}
J.~M.~Soares,
Nucl.\ Phys.\ B {\bf 367}, 575 (1991).

\bibitem{Kagan:1998bh}
A.~L.~Kagan and M.~Neubert,
Phys.\ Rev.\ D {\bf 58}, 094012 (1998).


\bibitem{delAguila:1986ne}
F.~del Aguila, M.~K.~Chase and J.~Cortes,
Nucl.\ Phys.\ B {\bf 271}, 61 (1986);
G.~C.~Branco and L.~Lavoura,
Nucl.\ Phys.\ B {\bf 278}, 738 (1986);
P.~Langacker and D.~London,
Phys.\ Rev.\ D {\bf 38}, 886 (1988);
Y.~Nir and D.~J.~Silverman,
Phys.\ Rev.\ D {\bf 42}, 1477 (1990);
D.~Silverman,
Phys.\ Rev.\ D {\bf 45}, 1800 (1992).

\bibitem{Branco:1993wr}
G.~C.~Branco, T.~Morozumi, P.~A.~Parada and M.~N.~Rebelo,
Phys.\ Rev.\ D {\bf 48}, 1167 (1993).

\bibitem{Barger:1995dd}
V.~Barger, M.~S.~Berger and R.~J.~Phillips,
Phys.\ Rev.\ D {\bf 52}, 1663 (1995).

\bibitem{Gronau:1997rv}
M.~Gronau and D.~London,
Phys.\ Rev.\ D {\bf 55}, 2845 (1997).

\bibitem{Barenboim:2001fd}
G.~Barenboim, F.~J.~Botella and O.~Vives,
Nucl.\ Phys.\ B {\bf 613}, 285 (2001).

\bibitem{Handoko:1995xw}
L.~T.~Handoko and T.~Morozumi,
Mod.\ Phys.\ Lett.\ A {\bf 10}, 309 (1995)
[Erratum-ibid.\ A {\bf 10}, 1733 (1995)].

\bibitem{Bhattacharyya:1994vp}
G.~Bhattacharyya, G.~C.~Branco and D.~Choudhury,
Phys.\ Lett.\ B {\bf 336}, 487 (1994)
[Erratum-ibid.\ B {\bf 340}, 266 (1994)].

\bibitem{Aoki:2000ze}
M.~Aoki, E.~Asakawa, M.~Nagashima, N.~Oshimo and A.~Sugamoto,
Phys.\ Lett.\ B {\bf 487}, 321 (2000).

\bibitem{Chang:2000gz}
C.~V.~Chang, D.~Chang and W.~Keung,
Phys.\ Rev.\ D {\bf 61}, 053007 (2000)
and contribution to BCP3, Taipei, Taiwan, December 1999.

\bibitem{vectorquarkpapers}
L.~T.~Handoko,
Nuovo Cim.\ A {\bf 111}, 95 (1998);
%
G.~Barenboim, F.~J.~Botella and O.~Vives,
Phys.\ Rev.\ D {\bf 64}, 015007 (2001);
%
Y.~Liao and X.~Li,
Phys.\ Lett.\ B {\bf 503}, 301 (2001);
%
M.~Aoki, G.~Cho, M.~Nagashima and N.~Oshimo,
hep-ph/0102165;
%
M.~R.~Ahmady, M.~Nagashima and A.~Sugamoto,
Phys.\ Rev.\ D {\bf 64}, 054011 (2001).

\bibitem{Atwood:1997zr}
D.~Atwood, M.~Gronau and A.~Soni,
Phys.\ Rev.\ Lett.\  {\bf 79}, 185 (1997).

\bibitem{Gabbiani:1996hi}
F.~Gabbiani, E.~Gabrielli, A.~Masiero and L.~Silvestrini,
Nucl.\ Phys.\ B {\bf 477}, 321 (1996).

\bibitem{Arhrib:2001gr}
A.~Arhrib, C.~K.~Chua and W.~S.~Hou,
Eur.\ Phys.\ J.\ C {\bf 21}, 567 (2001).

\bibitem{alex} J. Alexander, talk given at BCP4, Ise, Japan, February 2001,\\
   http://www.hepl.phys.nagoya-u.ac.jp/public/bcp4/presentation/23am/alexander.pdf.

\bibitem{Abe:2001qh}
K.~Abe {\it et al.}  [Belle Collaboration],
hep-ex/0107072.

\bibitem{Hurth:2001yb}
T.~Hurth and T.~Mannel,
Phys.\ Lett.\ B {\bf 511}, 196 (2001);
T.~Hurth and T.~Mannel, hep-ph/0109041.

\bibitem{newakeroydreck}
A.~G.~Akeroyd and S.~Recksiegel, work in progress.

\end{thebibliography}
\end{document}